\documentclass[twocolumn]{revtex4-1}
\usepackage{amsmath}
\usepackage{bbold}
\usepackage{graphicx}
\usepackage{epstopdf}
\usepackage{braket}

\graphicspath{{./graphicsInNotes/}}
\begin{document}
\title{Sensing Atomic Motion from the Zero Point to Room Temperature with Ultrafast Atom Interferometry}
\date{\today}
\author{K. G. Johnson}
\author{B. Neyenhuis}
\author{J. Mizrahi}
\author{J. D. Wong-Campos}
\author{C. Monroe}
\affiliation{Joint Quantum Institute, University of Maryland Department of Physics and National Institute of Standards and Technology, College Park, Maryland 20742, USA}
\begin{abstract}
We sense the motion of a trapped atomic ion using a sequence of state-dependent ultrafast momentum kicks.  We use this atom interferometer to characterize a nearly-pure quantum state with $n=1$ phonon and accurately measure thermal states ranging from near the zero-point energy to \(\bar{n}\sim 10^4\), with the possibility of extending at least 100 times higher in energy. The complete energy range of this method spans from the ground state to far outside of the Lamb-Dicke regime, where atomic motion is greater than the optical wavelength. Apart from thermometry, these interferometric techniques are useful for characterizing ultrafast entangling gates between multiple trapped ions.
\end{abstract}
\maketitle


There is great interest in the ultrafast quantum control of trapped ions, including the preparation of nonclassical states of motion \cite{Poyatos_96_PRA}, entangling quantum logic gates \cite{Duan_04_PRL, Garcia-Ripoll_03_PRL}, and ground state cooling on time-scales shorter than the period of ion harmonic motion \cite{Machnes_10_PRL}.  Experimental demonstrations of ultrafast trapped ion control include single qubit rotations \cite{Campbell_10_PRL,Ospelkaus_11_Nature} and spin-motion entanglement \cite{Mizrahi_13_PRL,Mizrahi_13_APB}. These operations can be many orders of magnitude faster than conventional techniques, less sensitive to noise, and scalable to large numbers of ions where the couplings occur through local modes of motion \cite{Duan_04_PRL}.  Moreover, these ultrafast techniques work far outside the Lamb-Dicke regime, where the extent of the atomic motion is greater than the optical wavelength \cite{Leibfried_03_RMP}. 

Ultrafast sensing of atomic motion allows measurements over a wide range of energies, from the zero-point (average phonon ocupation number $\bar{n}=0)$ to potentially above room-temperature ($\bar{n} \sim 10^6$ for typical ion traps). Ultrafast partial state tomography (defined later in this letter) on thermal states improves upon the dynamic range achieved with thermometry using dark resonances \cite{Rossnagel_2015_NJP}. It also complements conventional methods of thermometry, including measurements of the motionally-induced upper and lower sideband asymmetries \cite{Turchette_2000_PRA} and the thermal suppression of induced transitions (Debye-Waller factors) \cite{Roos_thesis}.  However, both of these other methods break down when the atomic motion is outside of the Lamb-Dicke regime, typically around \(\bar{n}>10\).  Measuring the entire Doppler-broadened envelope of sidebands \cite{Shu_14_PRA} provides a more general measurement of thermal states, but can be difficult due to the bandwidth required to excite multiple sidebands.  Here we use ultrafast techniques for accurate thermometry of ion motion ranging from $\bar{n}\sim 0.1$ to $\bar{n}\sim 10^4$ and show how this method extends to higher energies. We also measure particular quantum states through more complete motional tomography.   

In this experiment, we trap a $^{171}$Yb$^+$ ion in a linear radio frequency Paul trap and probe the motion along a single radial mode of motion with secular trap frequency \(\omega_{t}/2\pi \approx 1 \) MHz, as described in Ref. \cite{Mizrahi_13_PRL}.  
The $\ket{F=0,m_{f}=0} \equiv \ket{\downarrow}$ and $\ket{F=1,m_{f}=0}\equiv \ket{\uparrow}$ hyperfine levels of the \(^{2}S_{1/2}\) electronic ground state are used as the qubit, or effective spin, and are separated by the splitting \(\omega_{0}/2\pi=12.642815\) GHz. The ion is laser-cooled to near the Doppler limit using the 
$^{2}S_{1/2}$ to $^{2}P_{1/2}$ transition at a wavelength of $369.5$ nm, and optically pumped via the \(^{2}P_{1/2}\), \(F=1\) levels to the $\ket{\downarrow}$ level during initial state preparation.  Qubit state detection is performed by collecting state-dependent fluorescence \cite{Olmschenk_07_PRA}. The qubit state in these experiments is detected with an efficiency above $0.997$ using an imaging objective with 0.6 numerical aperture and a photomultiplier tube \cite{Noek_2013_OL}.

We create a spin-dependent kick (SDK) by shaping individual pulses extracted from a mode-locked laser with center wavelength \(2\pi/k \approx 355\) nm, pulse duration \(\tau \sim 10\) ps, and repetition rate of \(f_{rep} = 118\) MHz. A series of optical delay paths in the form of three sequential Mach-Zehnder interferometers shape the pulse, dividing it into eight sub-pulses that are spaced to collectively flip the qubit spin while generating spin-dependence in the momentum transfer \cite{Mizrahi_13_PRL}. In addition to setting proper delay lengths, we adjust the pulse energy to give a complete spin flip \cite{Mizrahi_13_APB}. The pulse is applied to the ion in a counter-propagating geometry, ideally creating the evolution operator \(\hat{U}_{SDK}=\hat{\mathcal{D}}[i \eta]\hat{\sigma}_{+}+\hat{\mathcal{D}}[-i \eta]\hat{\sigma}_{-}\) \footnote{In this evolution operator, we suppress the optical difference phase between the two beams generating the SDK because it is constant over the time scale of one experiment and ultimately cancels.}, where $\hat{\sigma}_{\pm}$ are the qubit raising and lowering operators. The displacement operator \(\hat{\mathcal{D}}[\pm i\eta]\) imparts momentum $\Delta p = \pm 2 \hbar k = \pm \eta p_0$, where \(p_0 = \sqrt{2M\hbar\omega_t}\) is the zero-point spread of momentum in the trap, $M$ is the atomic mass, and $\eta=0.2$ is the Lamb-Dicke parameter associated with this momentum transfer. The impulsive SDK operation occurs on a time scale much faster than the trap period (\(\tau \ll 1/\omega_t\)), and the spin population transfer from $\ket{\downarrow}$ to $\ket{\uparrow}$ is measured to have a fidelity of $0.993(2)$ \cite{Mizrahi_13_APB}. Because each SDK operation provides a momentum kick and flips the spin, immediately applying a second SDK would simply undo the first.  However, by waiting one half of the trap period between SDKs, we can concatenate $N$ individual kicks to create a larger effective SDK with $\Delta p = \pm 2 N \hbar k = \pm N \eta p_0$.

\begin{figure}
	\includegraphics[scale=0.58]{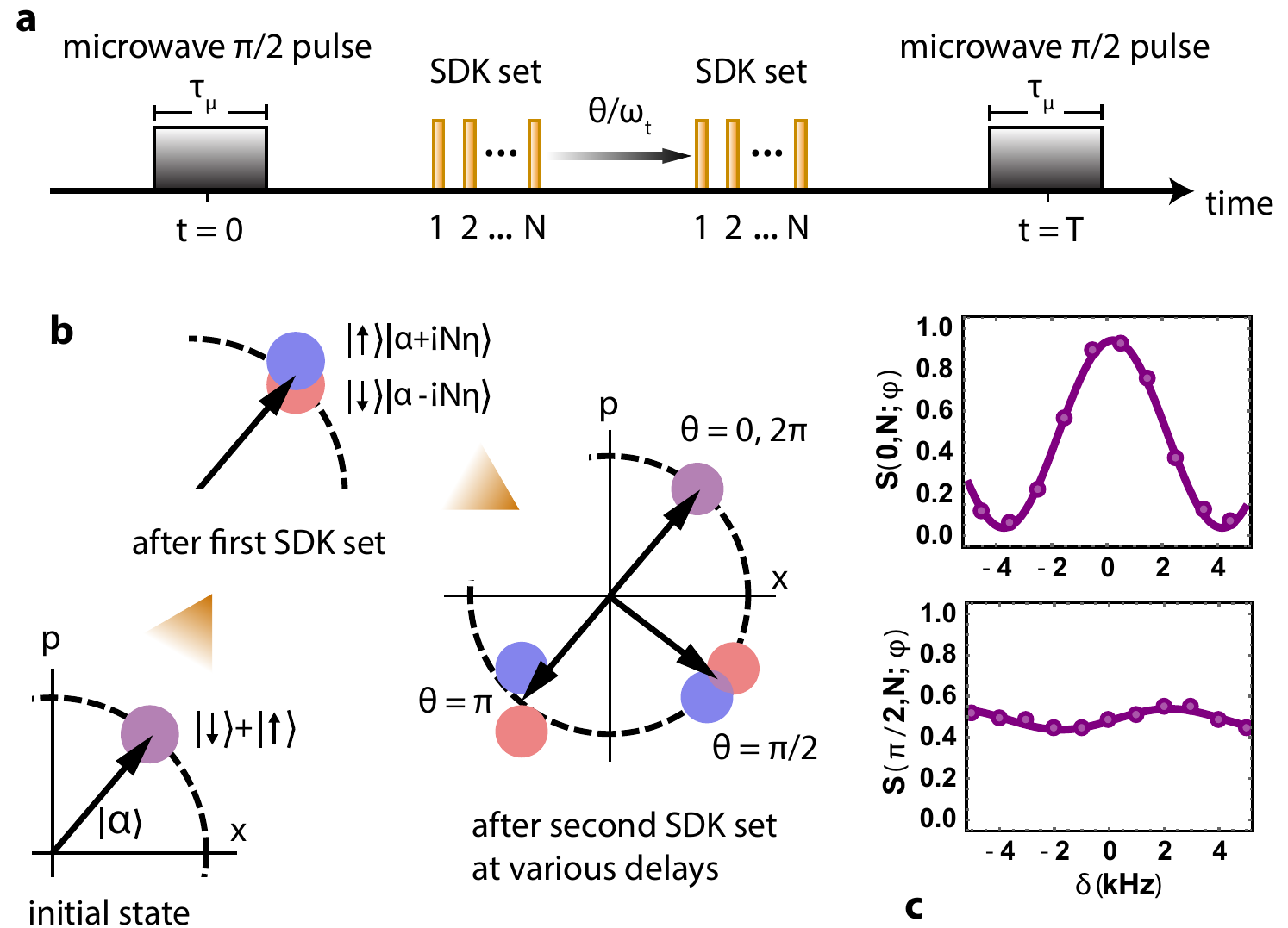}
	\caption{\label{C_20_error_and_Ramsey.pdf} (a) Timeline of a single experiment, where a full SDK set is made of $N$ single SDKs. (b) Phase space diagram of an initial state $(\ket{\downarrow}+\ket{\uparrow})\ket{\alpha}$ evolving under two sets of SDKs separated by time delay $\theta/\omega_t$, where $\ket{\alpha}$ is a coherent state of motion. (c) Typical Ramsey fringes as a function of microwave frequency detuning \(\delta\). These two plots correspond to the points \(\theta=0\) and \(\theta=\pi/2\) of an initial thermal state ($N=1$ for the data shown).}
\end{figure}

 Techniques dealing with Ramsey spectroscopy on states coherently displaced by spin dependent forces have been demonstrated in creating Schrodinger cat states \cite{Monroe_96_Science} and measuring spin dephasing in 2D ion crystals \cite{Sawyer_2014_PRA}. In this experiment, we create an interferometer to sense motion by applying two sets of $N$ SDK operations within a Ramsey experiment on the qubit levels with time duration $T$. First the ion is prepared in a coherent superposition of $\ket{\downarrow}$ and $\ket{\uparrow}$ by applying a near-resonant microwave $\pi/2$ pulse of duration $\tau_{\mu}$. Following the first set of $N$ SDKs, the ion evolves for a time \(\theta/\omega_t\) and then a second set of $N$ SDKs is applied. Finally, another $\pi/2$ pulse with the same duration and tuning drives the qubit to close the Ramsey interferometer. This sequence is diagrammed in Fig. \ref{C_20_error_and_Ramsey.pdf}a. By scanning the microwave detuning $\delta \ll 1/\tau_\mu$ from resonance, we observe sets of Ramsey fringes with phase $\phi = \delta T$ that chronicle the ion motion (shown in Fig. \ref{C_20_error_and_Ramsey.pdf}b and \ref{C_20_error_and_Ramsey.pdf}c).

For a pure initial state \(\ket{\Psi^{\alpha}}_i=\ket{\downarrow}\ket{\alpha}\), where \(\alpha\) is a coherent state of the ion motion, the state following the Ramsey experiment is
\begin{align}
\ket{\Psi^\alpha}=\frac{1}{2} {}& [e^{i\gamma}(\ket{\downarrow}+i e^{-i \phi}\ket{\uparrow})\ket{(\alpha+i N\eta)e^{-i \theta}-i N\eta}+ \nonumber\\& i e^{-i\gamma}(\ket{\uparrow}+i e^{i \phi}\ket{\downarrow})\ket{(\alpha-i N\eta)e^{-i \theta}+i N\eta}],
\end{align} 
where $\gamma=N\eta[\text{Re}(\alpha)(1-\cos\theta)-\text{Im}(\alpha) \sin\theta]$.

Given an arbitrary initial state of motion in phase space described by the Glauber P-distribution \cite{Glauber_1963_PR,Sudarshan_1963_PRL}, the final density matrix is \(\hat{\rho} = \int P(\alpha) \ket{\Psi^\alpha}\bra{\Psi^\alpha} d^2\alpha\). The probability of measuring the state spin-up after the Ramsey experiment is therefore
\begin{align}
S {}& (\theta,N;\phi) = \bra{\uparrow}\hat{\rho}\ket{\uparrow}
\nonumber\\& =\frac{1}{2} + \frac{1}{2} \int P(\alpha) e^{-4(N\eta)^{2}(1-\cos\theta)}\cos(4\gamma-\phi) d^2 \alpha. \label{Brightness}
\end{align}

Two types of motional state that are readily accessible in the laboratory are thermal states and small Fock states. First we discuss ultrafast partial state tomography to determine the average phonon number in a thermal state. Then we extend this method to create a nearly complete map of the motion of an $n=1$ Fock state in phase space, showing clear nonclassical signatures.

For an ion prepared in a thermal state with mean phonon number $\bar{n}$ and P-function \(P_{therm}(\alpha)=\frac{1}{\pi \bar{n}}e^{-|\alpha|^{2}/\bar{n}}\), Eq. \ref{Brightness} yields an expected Ramsey fringe pattern
\begin{equation}
S_{therm}(\theta,N;\phi) = \frac{1}{2}+\frac{1}{2}e^{-4(N\eta)^{2}(2\bar{n}+1)(1-cos\theta)} cos\phi \label{contrast}.
\end{equation}
The fringe contrast has periodic revival peaks at $\theta=2\pi m$, where $m$ is a positive integer.  For a hot ion where $\bar{n} \gg 1/(N\eta)^2$, these revivals in contrast become narrow and approximately Gaussian with full width at half maximum FWHM$= 0.83/(N \eta \sqrt{\bar{n}})$. With $N=1$, we measure the Ramsey fringe contrast as a function of $\theta$ for a variety of initial thermal states of motion, and fit the contrast revival peaks to Eq. 3 to determine the average phonon number $\bar{n}$ of the thermal state \cite{Turchette_00_PRA,Myatt_00_Nature,Poshinger_2010_PRL}. In the fit, we allow the peak Ramsey contrast at $\theta = 2\pi m$ to be less than unity in order to parametrize imperfect fidelity of the SDK operations. This reduction in fidelity is mainly attributed to variations in the Raman beam intensity over the spatial extent of the ion wave packet, and becomes apparent at high $\bar{n}$. This does not significantly affect the width of the contrast revival peak or the accuracy of the thermometer, and can be mended by widening the beam waist.
\begin{figure}
	\includegraphics[scale=0.4]{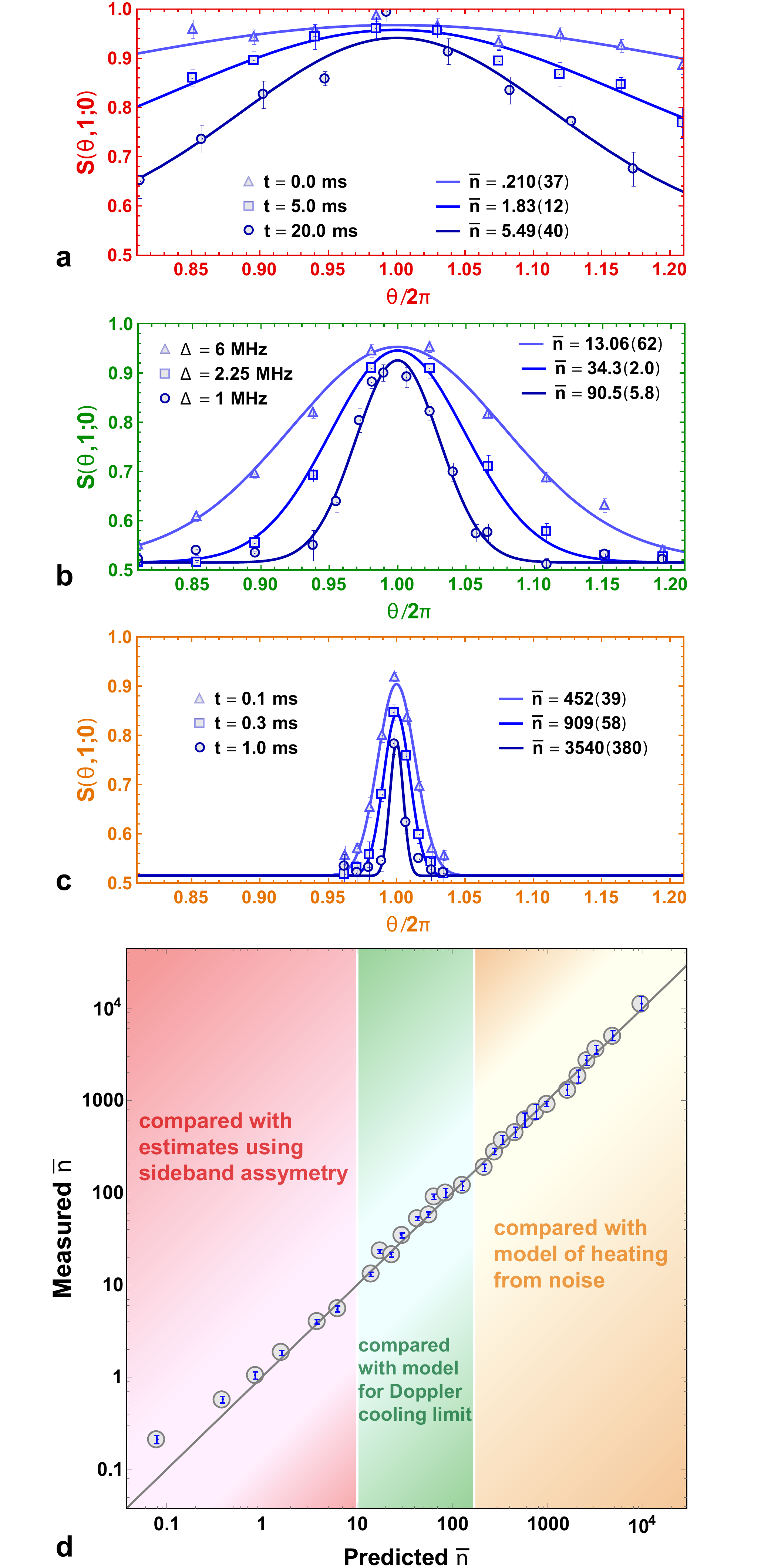}
	\caption{\label{complete_temp_plots.pdf} Ultrafast sensing measurements of $\bar{n}$ (with $N=1$): (a) Sampling of Ramsey revival contrast lineshapes with initial states prepared by resolved sideband cooling to the ground state and subsequent heating. Data is fit to $S_{therm}(\theta,1;\phi)$. The amplitude of each fit is a free parameter to account for SDK infidelity (also done in \textbf{b} and \textbf{c}). This does not significantly affect the width of the peak, which is used to determine \(\bar{n}\). (b) Sampling of Ramsey revival contrast lineshapes with initial states prepared by Doppler cooling only, with $\bar{n}$ varied by changing the cooling beam detuning. (c) Sampling of Ramsey revival contrast lineshapes with initial states prepared by inducing a high heating rate with white noise applied to a trap electrode. (d) Measurements of \(\bar{n}\) versus predicted values. There are three regimes of thermal state preparation--red being sideband-cooling then heating (see (a)), green being Doppler cooling with different detunings (see (b)) and orange being heating with applied noise (see (c)).}
\end{figure}

Ramsey contrast revival lineshapes are measured in experiments spanning over five orders of magnitude in $\bar{n}$, as shown in Figs. 2a-c. Figure \ref{complete_temp_plots.pdf}d shows these measurements plotted versus the expected value of $\bar{n}$ from theory and other measurements.  The figure is broken into three regions according to the manner in which the motional state is prepared and calibrated before measurement of the contrast revival lineshapes. Low energy thermal states ($\bar{n}<10$) are generated by first sideband-cooling the ion to its zero point motion and then allowing the ion to weakly heat in the trap by known amounts. In this regime, we compare ultrafast interferometric measurements of $\bar{n}$ (shown in Fig. \ref{complete_temp_plots.pdf}a) to values extracted from measured sideband asymmetries \cite{Wineland_1998_bible}. The deviation of the two measurements are shown in the red section of Fig. \ref{complete_temp_plots.pdf}d.

For thermal states $10\lesssim \bar{n} \lesssim 150$, the ion is prepared by Doppler cooling with various frequency detunings from resonance. Ultrafast measurements in this regime are shown in Fig. \ref{complete_temp_plots.pdf}b. Each of these measurements is compared to the predicted value of $\bar{n}$ from Doppler cooling theory \cite{Wineland_1979_PRA} and plotted against each other in the green section of Fig. \ref{complete_temp_plots.pdf}d.  As a check on the expected values of $\bar{n}$ in this range, we also measure the Debye-Waller suppression of Rabi flopping transitions between the ion qubit states \cite{Wineland_1998_bible} for several cases, resulting in expected values consistent with Doppler theory.

Hot thermal states are prepared by inducing a high heating rate with a noisy electrical potential to a trap electrode for varied amounts of time after Doppler cooling. The ultrafast measurements of these states are shown in Fig. \ref{complete_temp_plots.pdf}c. Measurements in this regime are compared to a predicted $\bar{n}$ given by the equation $\dot{\bar{n}}=\frac{e^2 S_V(\omega_t)}{4 M \hbar \omega_t d^2}$ \cite{Turchette_2000_PRA}, where $e$ is the ion charge, and $S_V(\omega_t)$ ($\textrm{V}^2/\textrm{Hz}$) is the applied power noise spectral density of the electric-potential, which is white over the measurement bandwidth. The effective distance $d$ of the electrode to the ion  is calibrated by applying a static potential offset to the same electrode and observing the resulting displacement of the ion in space. The predicted and measured values for this regime are plotted against each other in the orange region of Fig. \ref{complete_temp_plots.pdf}d.

\begin{figure*}
 \includegraphics[scale=0.58]{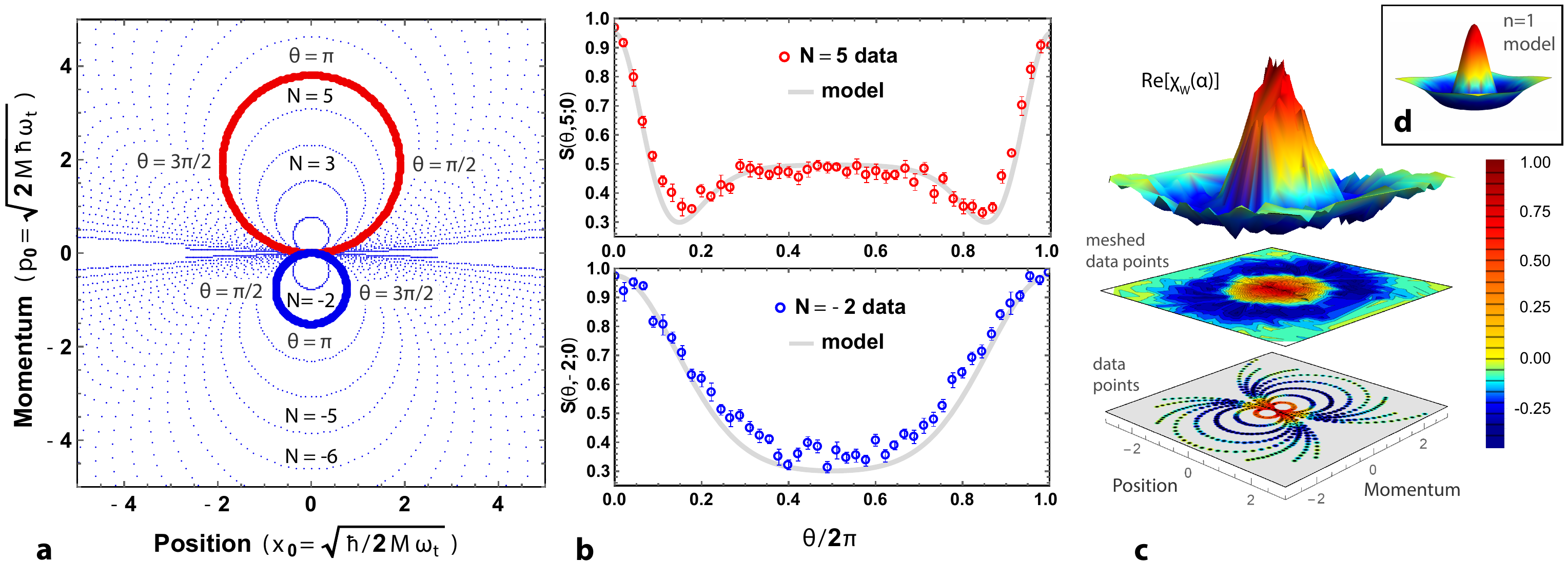}
 \caption{\label{fock_1_PlotsAndPhaseSpace.pdf}(a) Points in phase space accessible in our tomographic measurements.  The radius of each circle ($2N\eta$) is set by the number of kicks $N$, and the angular position on each circle is set by the SDK delay $\theta$. The sign of $N$ represents the direction of the initial momentum kicks associated with the spin flip operators. (b) Sample of measurements of the Ramsey fringe at $\phi=0$ for a nominal $n=1$ Fock state, using two sets of kicks with $N=5$ (red) and $N=-2$ (blue) and scanning the delay $\theta$. The coordinates of these particular scans in phase space are highlighed in \textbf{a}. (c) Motional state tomography of an ion prepared in the $n=1$ Fock state. In ascending order: the value-colored data points of \(\text{Re}[\chi_{W}(\alpha)]\) taken on 16 rings in phase space set by $\pm N$ where $N=1,2,3,4,5,6,8,10$, the interpolated data mapped in contour, and a 3D interpolation of the data. (d) Theory prediction for a Fock state with \(n=1\).}
\end{figure*}

We next perform more complete tomography of a nearly pure quantum state of motion by extracting the characteristic function
\begin{equation}
\chi_{W}(\alpha) = e^{-|\alpha|^2/2} \int P(\beta) e^{2i \text{Im}(\alpha \beta^*)} d^2\beta. \label{char_transform}
\end{equation}
This quasiprobability distribution contains all the information about the quantum state and is the Fourier transform of the better-known Wigner distribution \cite{Gehrke_thesis,Wigner_1932_PR}.

In terms of the observable $S(\theta,N;\phi)$, $\chi_{W}(\alpha)$ is given by
\begin{align}
\text{Re}[\chi_W(\alpha)] {}& =2 S(\theta,N;0) - 1
\\\text{Im}[\chi_W(\alpha)] \nonumber& =2 S(\theta,N;\frac{\pi}{2}) - 1,
\label{char_function}
\end{align}
where $\alpha = 2N\eta[\sin\theta+i(1-\cos\theta)]$. Scanning \(\theta\) and $N$ while measuring \(S(\theta,N;\phi)\) maps the characteristic function over rings in phase space, shown in Fig. \ref{fock_1_PlotsAndPhaseSpace.pdf}a. In order to scan the negative imaginary part of $\alpha$, we can change the direction of the initial momentum kick associated with the spin flip operators by shifting the relative optical phase of the counter-propagating beams by $\pi$ \cite{Mizrahi_13_PRL}. These reversed kicks can be thought of as effectively flipping the sign of $\eta$, and for simplicity, we represent them here by negative values of $N$.

We measure the characteristic function \(\chi_{W}(\alpha)\) of the ion in the n=1 Fock state, prepared by sideband cooling to the ground state and transferring population to the n=1 state through application of a blue sideband operation \cite{Leibfried_03_RMP}. To have a grid that spans the domain of the state, we scan around 16 rings in phase space set by $\pm N$, where $N=1, 2, 3, 4, 5, 6, 8, 10$.  Two of the 16 rings along which we measure are highlighted in Fig. \ref{fock_1_PlotsAndPhaseSpace.pdf}a, and plots of \(S(\theta,N;0)\) versus $\theta$ along those two rings are shown in Fig. \ref{fock_1_PlotsAndPhaseSpace.pdf}b. Notice in Fig. \ref{fock_1_PlotsAndPhaseSpace.pdf}b that the larger SDK set ($N=5$) separates the interferometer enough to see the oscillation of the motional distribution, while the smaller SDK set does not. Mapping along all 16 curves gives a nearly complete motional state map. The real part of the characteristic function is shown in Fig. \ref{fock_1_PlotsAndPhaseSpace.pdf}c alongside the corresponding model of \(\text{Re}[\chi_{W}(\alpha)]\) for an $n=1$ Fock state in Fig. \ref{fock_1_PlotsAndPhaseSpace.pdf}d. The negative values of the characteristic quasiprobability function highlight the nonclassical nature of the motional state of the ion.

These ultrafast tomographic techniques are capable of measuring motional energies far beyond the data presented here, which was limited to $\bar{n} \sim 10^4$ because of re-cooling issues during state preparation. In the experiment, we scan the interferometric angular delay $\theta$ in steps set by the repetition rate of the laser, giving a resolution of $\omega_t/f_{rep} \sim 50$ mrad.  For lineshapes narrower than this laser repetition rate limit, we scan $\theta$ by changing the trap frequency $\omega_{t}$ though accurate control of the trap rf drive voltage. With fine drive voltage control, we can achieve a resolution in $\theta$ of $0.1$ mrad, which would correspond to a Ramsey revival linewidth from a thermal state with $\bar{n} \sim 10^9$.  Other factors also come into play when measuring such high-energy states: First, the spatial extent of motion swells beyond the laser beam waist. At \(\bar{n}=10^6\) for instance, or equivalent temperature \(T=\hbar\omega_t \bar{n}/k_B=80\)K, the ion would experience a significant gradient in the Rabi frequency across a beam with a 3\(\mu\)m waist. A second factor is the decreased detection fluorescence due to larger Doppler shifts at these energies. The detection fluorescence at $\bar{n}=10^6$ would be reduced by a factor of \(\sim 10^3\) from a cold ion \cite{Wineland_1979_PRA}. Finally, when measuring these very narrow lineshapes, instabilities in the trap frequency $\omega_t$ and laser repetition rate \(f_{rep}\) would have to be sufficiently stable over the measurement time.  At \(\bar{n}=10^6\), this would require a fractional stability from both the trap frequency and laser repetition rate of better than $0.1\%$. These factors put ultrafast interferometric measurements of $\bar{n}\ge 10^6$ neither fundamentally nor technically beyond reach.
\begin{acknowledgments}
	This work is supported by the Army Research Office and NSF Physics Frontier Center at JQI.
\end{acknowledgments}
\bibliography{Kale_Johnson_Ultrafast}

\begin{thebibliography}{27}%
\makeatletter
\providecommand \@ifxundefined [1]{%
 \@ifx{#1\undefined}
}%
\providecommand \@ifnum [1]{%
 \ifnum #1\expandafter \@firstoftwo
 \else \expandafter \@secondoftwo
 \fi
}%
\providecommand \@ifx [1]{%
 \ifx #1\expandafter \@firstoftwo
 \else \expandafter \@secondoftwo
 \fi
}%
\providecommand \natexlab [1]{#1}%
\providecommand \enquote  [1]{``#1''}%
\providecommand \bibnamefont  [1]{#1}%
\providecommand \bibfnamefont [1]{#1}%
\providecommand \citenamefont [1]{#1}%
\providecommand \href@noop [0]{\@secondoftwo}%
\providecommand \href [0]{\begingroup \@sanitize@url \@href}%
\providecommand \@href[1]{\@@startlink{#1}\@@href}%
\providecommand \@@href[1]{\endgroup#1\@@endlink}%
\providecommand \@sanitize@url [0]{\catcode `\\12\catcode `\$12\catcode
  `\&12\catcode `\#12\catcode `\^12\catcode `\_12\catcode `\%12\relax}%
\providecommand \@@startlink[1]{}%
\providecommand \@@endlink[0]{}%
\providecommand \url  [0]{\begingroup\@sanitize@url \@url }%
\providecommand \@url [1]{\endgroup\@href {#1}{\urlprefix }}%
\providecommand \urlprefix  [0]{URL }%
\providecommand \Eprint [0]{\href }%
\providecommand \doibase [0]{http://dx.doi.org/}%
\providecommand \selectlanguage [0]{\@gobble}%
\providecommand \bibinfo  [0]{\@secondoftwo}%
\providecommand \bibfield  [0]{\@secondoftwo}%
\providecommand \translation [1]{[#1]}%
\providecommand \BibitemOpen [0]{}%
\providecommand \bibitemStop [0]{}%
\providecommand \bibitemNoStop [0]{.\EOS\space}%
\providecommand \EOS [0]{\spacefactor3000\relax}%
\providecommand \BibitemShut  [1]{\csname bibitem#1\endcsname}%
\let\auto@bib@innerbib\@empty
\bibitem [{\citenamefont {Poyatos}\ \emph {et~al.}(1996)\citenamefont
  {Poyatos}, \citenamefont {Cirac}, \citenamefont {Blatt},\ and\ \citenamefont
  {Zoller}}]{Poyatos_96_PRA}%
  \BibitemOpen
  \bibfield  {author} {\bibinfo {author} {\bibfnamefont {J.~F.}\ \bibnamefont
  {Poyatos}}, \bibinfo {author} {\bibfnamefont {J.~I.}\ \bibnamefont {Cirac}},
  \bibinfo {author} {\bibfnamefont {R.}~\bibnamefont {Blatt}}, \ and\ \bibinfo
  {author} {\bibfnamefont {P.}~\bibnamefont {Zoller}},\ }\href@noop {}
  {\bibfield  {journal} {\bibinfo  {journal} {Phys. Rev. A}\ }\textbf {\bibinfo
  {volume} {54}},\ \bibinfo {pages} {1532} (\bibinfo {year}
  {1996})}\BibitemShut {NoStop}%
\bibitem [{\citenamefont {Duan}(2004)}]{Duan_04_PRL}%
  \BibitemOpen
  \bibfield  {author} {\bibinfo {author} {\bibfnamefont {L.-M.}\ \bibnamefont
  {Duan}},\ }\href@noop {} {\bibfield  {journal} {\bibinfo  {journal} {Phys.
  Rev. Lett.}\ }\textbf {\bibinfo {volume} {93}},\ \bibinfo {pages} {100502}
  (\bibinfo {year} {2004})}\BibitemShut {NoStop}%
\bibitem [{\citenamefont {Garcia-Ripoll}\ \emph {et~al.}(2003)\citenamefont
  {Garcia-Ripoll}, \citenamefont {Zoller},\ and\ \citenamefont
  {Cirac}}]{Garcia-Ripoll_03_PRL}%
  \BibitemOpen
  \bibfield  {author} {\bibinfo {author} {\bibfnamefont {J.~J.}\ \bibnamefont
  {Garcia-Ripoll}}, \bibinfo {author} {\bibfnamefont {P.}~\bibnamefont
  {Zoller}}, \ and\ \bibinfo {author} {\bibfnamefont {J.~I.}\ \bibnamefont
  {Cirac}},\ }\href@noop {} {\bibfield  {journal} {\bibinfo  {journal} {Phys.
  Rev. Lett.}\ }\textbf {\bibinfo {volume} {91}},\ \bibinfo {pages} {157901}
  (\bibinfo {year} {2003})}\BibitemShut {NoStop}%
\bibitem [{\citenamefont {Machnes}\ \emph {et~al.}(2010)\citenamefont
  {Machnes}, \citenamefont {Plenio}, \citenamefont {Reznik}, \citenamefont
  {Steane},\ and\ \citenamefont {Retzker}}]{Machnes_10_PRL}%
  \BibitemOpen
  \bibfield  {author} {\bibinfo {author} {\bibfnamefont {S.}~\bibnamefont
  {Machnes}}, \bibinfo {author} {\bibfnamefont {M.~B.}\ \bibnamefont {Plenio}},
  \bibinfo {author} {\bibfnamefont {B.}~\bibnamefont {Reznik}}, \bibinfo
  {author} {\bibfnamefont {A.~M.}\ \bibnamefont {Steane}}, \ and\ \bibinfo
  {author} {\bibfnamefont {A.}~\bibnamefont {Retzker}},\ }\href@noop {}
  {\bibfield  {journal} {\bibinfo  {journal} {Phys. Rev. Lett.}\ }\textbf
  {\bibinfo {volume} {104}},\ \bibinfo {pages} {183001} (\bibinfo {year}
  {2010})}\BibitemShut {NoStop}%
\bibitem [{\citenamefont {Campbell}\ \emph {et~al.}(2010)\citenamefont
  {Campbell}, \citenamefont {Mizrahi}, \citenamefont {Quraishi}, \citenamefont
  {Senko}, \citenamefont {Hayes}, \citenamefont {Hucul}, \citenamefont
  {Matsukevich}, \citenamefont {Maunz},\ and\ \citenamefont
  {Monroe}}]{Campbell_10_PRL}%
  \BibitemOpen
  \bibfield  {author} {\bibinfo {author} {\bibfnamefont {W.~C.}\ \bibnamefont
  {Campbell}}, \bibinfo {author} {\bibfnamefont {J.}~\bibnamefont {Mizrahi}},
  \bibinfo {author} {\bibfnamefont {Q.}~\bibnamefont {Quraishi}}, \bibinfo
  {author} {\bibfnamefont {C.}~\bibnamefont {Senko}}, \bibinfo {author}
  {\bibfnamefont {D.}~\bibnamefont {Hayes}}, \bibinfo {author} {\bibfnamefont
  {D.}~\bibnamefont {Hucul}}, \bibinfo {author} {\bibfnamefont {D.~N.}\
  \bibnamefont {Matsukevich}}, \bibinfo {author} {\bibfnamefont
  {P.}~\bibnamefont {Maunz}}, \ and\ \bibinfo {author} {\bibfnamefont
  {C.}~\bibnamefont {Monroe}},\ }\href@noop {} {\bibfield  {journal} {\bibinfo
  {journal} {Phys. Rev. Lett.}\ }\textbf {\bibinfo {volume} {105}},\ \bibinfo
  {pages} {090502} (\bibinfo {year} {2010})}\BibitemShut {NoStop}%
\bibitem [{\citenamefont {Ospelkaus}\ \emph {et~al.}(2011)\citenamefont
  {Ospelkaus}, \citenamefont {Warring}, \citenamefont {Colombe}, \citenamefont
  {Brown}, \citenamefont {Amini}, \citenamefont {Leibfried},\ and\
  \citenamefont {Wineland}}]{Ospelkaus_11_Nature}%
  \BibitemOpen
  \bibfield  {author} {\bibinfo {author} {\bibfnamefont {C.}~\bibnamefont
  {Ospelkaus}}, \bibinfo {author} {\bibfnamefont {U.}~\bibnamefont {Warring}},
  \bibinfo {author} {\bibfnamefont {Y.}~\bibnamefont {Colombe}}, \bibinfo
  {author} {\bibfnamefont {K.~R.}\ \bibnamefont {Brown}}, \bibinfo {author}
  {\bibfnamefont {J.~M.}\ \bibnamefont {Amini}}, \bibinfo {author}
  {\bibfnamefont {D.}~\bibnamefont {Leibfried}}, \ and\ \bibinfo {author}
  {\bibfnamefont {D.~J.}\ \bibnamefont {Wineland}},\ }\href@noop {} {\bibfield
  {journal} {\bibinfo  {journal} {Nature}\ }\textbf {\bibinfo {volume} {476}},\
  \bibinfo {pages} {181} (\bibinfo {year} {2011})}\BibitemShut {NoStop}%
\bibitem [{\citenamefont {Mizrahi}\ \emph
  {et~al.}(2013{\natexlab{a}})\citenamefont {Mizrahi}, \citenamefont {Senko},
  \citenamefont {Campbell}, \citenamefont {Johnson}, \citenamefont {Conover},\
  and\ \citenamefont {Monroe}}]{Mizrahi_13_PRL}%
  \BibitemOpen
  \bibfield  {author} {\bibinfo {author} {\bibfnamefont {J.}~\bibnamefont
  {Mizrahi}}, \bibinfo {author} {\bibfnamefont {C.}~\bibnamefont {Senko}},
  \bibinfo {author} {\bibfnamefont {W.~C.}\ \bibnamefont {Campbell}}, \bibinfo
  {author} {\bibfnamefont {K.~G.}\ \bibnamefont {Johnson}}, \bibinfo {author}
  {\bibfnamefont {C.~W.~S.}\ \bibnamefont {Conover}}, \ and\ \bibinfo {author}
  {\bibfnamefont {C.}~\bibnamefont {Monroe}},\ }\href@noop {} {\bibfield
  {journal} {\bibinfo  {journal} {Phys. Rev. Lett.}\ }\textbf {\bibinfo
  {volume} {110}},\ \bibinfo {pages} {203011} (\bibinfo {year}
  {2013}{\natexlab{a}})}\BibitemShut {NoStop}%
\bibitem [{\citenamefont {Mizrahi}\ \emph
  {et~al.}(2013{\natexlab{b}})\citenamefont {Mizrahi}, \citenamefont
  {Neyenhuis}, \citenamefont {Johnson}, \citenamefont {Campbell}, \citenamefont
  {Senko}, \citenamefont {Hayes},\ and\ \citenamefont
  {Monroe}}]{Mizrahi_13_APB}%
  \BibitemOpen
  \bibfield  {author} {\bibinfo {author} {\bibfnamefont {J.}~\bibnamefont
  {Mizrahi}}, \bibinfo {author} {\bibnamefont {Neyenhuis}}, \bibinfo {author}
  {\bibfnamefont {K.~G.}\ \bibnamefont {Johnson}}, \bibinfo {author}
  {\bibfnamefont {W.~C.}\ \bibnamefont {Campbell}}, \bibinfo {author}
  {\bibfnamefont {C.}~\bibnamefont {Senko}}, \bibinfo {author} {\bibfnamefont
  {D.}~\bibnamefont {Hayes}}, \ and\ \bibinfo {author} {\bibfnamefont
  {C.}~\bibnamefont {Monroe}},\ }\href@noop {} {\bibfield  {journal} {\bibinfo
  {journal} {Appl. Phys. B}\ }\textbf {\bibinfo {volume} {114}},\ \bibinfo
  {pages} {45} (\bibinfo {year} {2013}{\natexlab{b}})}\BibitemShut {NoStop}%
\bibitem [{\citenamefont {Leibfried}\ \emph {et~al.}(2003)\citenamefont
  {Leibfried}, \citenamefont {Blatt}, \citenamefont {Monroe},\ and\
  \citenamefont {Wineland}}]{Leibfried_03_RMP}%
  \BibitemOpen
  \bibfield  {author} {\bibinfo {author} {\bibfnamefont {D.}~\bibnamefont
  {Leibfried}}, \bibinfo {author} {\bibfnamefont {R.}~\bibnamefont {Blatt}},
  \bibinfo {author} {\bibfnamefont {C.}~\bibnamefont {Monroe}}, \ and\ \bibinfo
  {author} {\bibfnamefont {D.}~\bibnamefont {Wineland}},\ }\href@noop {}
  {\bibfield  {journal} {\bibinfo  {journal} {Rev. Mod. Phys.}\ }\textbf
  {\bibinfo {volume} {75}},\ \bibinfo {pages} {281} (\bibinfo {year}
  {2003})}\BibitemShut {NoStop}%
\bibitem [{\citenamefont {Rossnagel}\ \emph {et~al.}(2015)\citenamefont
  {Rossnagel}, \citenamefont {Tolazzi}, \citenamefont {Schmidt-Kaler},\ and\
  \citenamefont {Singer}}]{Rossnagel_2015_NJP}%
  \BibitemOpen
  \bibfield  {author} {\bibinfo {author} {\bibfnamefont {J.}~\bibnamefont
  {Rossnagel}}, \bibinfo {author} {\bibfnamefont {K.~N.}\ \bibnamefont
  {Tolazzi}}, \bibinfo {author} {\bibfnamefont {F.}~\bibnamefont
  {Schmidt-Kaler}}, \ and\ \bibinfo {author} {\bibfnamefont {K.}~\bibnamefont
  {Singer}},\ }\href@noop {} {\bibfield  {journal} {\bibinfo  {journal} {New J.
  Phys.}\ }\textbf {\bibinfo {volume} {17}},\ \bibinfo {pages} {045004}
  (\bibinfo {year} {2015})}\BibitemShut {NoStop}%
\bibitem [{\citenamefont {Turchette}\ \emph
  {et~al.}(2000{\natexlab{a}})\citenamefont {Turchette}, \citenamefont
  {Kielpinski}, \citenamefont {King}, \citenamefont {Myatt}, \citenamefont
  {Sachett}, \citenamefont {Itano}, \citenamefont {Monroe},\ and\ \citenamefont
  {Wineland}}]{Turchette_2000_PRA}%
  \BibitemOpen
  \bibfield  {author} {\bibinfo {author} {\bibfnamefont {Q.}~\bibnamefont
  {Turchette}}, \bibinfo {author} {\bibfnamefont {D.}~\bibnamefont
  {Kielpinski}}, \bibinfo {author} {\bibfnamefont {B.}~\bibnamefont {King}},
  \bibinfo {author} {\bibfnamefont {C.}~\bibnamefont {Myatt}}, \bibinfo
  {author} {\bibfnamefont {C.}~\bibnamefont {Sachett}}, \bibinfo {author}
  {\bibfnamefont {W.}~\bibnamefont {Itano}}, \bibinfo {author} {\bibfnamefont
  {C.}~\bibnamefont {Monroe}}, \ and\ \bibinfo {author} {\bibfnamefont {D.~J.}\
  \bibnamefont {Wineland}},\ }\href@noop {} {\bibfield  {journal} {\bibinfo
  {journal} {Phys. Rev. A}\ }\textbf {\bibinfo {volume} {61}},\ \bibinfo
  {pages} {063418} (\bibinfo {year} {2000}{\natexlab{a}})}\BibitemShut
  {NoStop}%
\bibitem [{\citenamefont {Roos}(2000)}]{Roos_thesis}%
  \BibitemOpen
  \bibfield  {author} {\bibinfo {author} {\bibfnamefont {C.}~\bibnamefont
  {Roos}},\ }\href@noop {} {Ph.D. thesis},\ \bibinfo  {school} {University of
  Innsbruck} (\bibinfo {year} {2000})\BibitemShut {NoStop}%
\bibitem [{\citenamefont {Shu}\ \emph {et~al.}(2014)\citenamefont {Shu},
  \citenamefont {Vittorini}, \citenamefont {Buikema}, \citenamefont {Nichols},
  \citenamefont {Volin}, \citenamefont {Stick},\ and\ \citenamefont
  {Brown}}]{Shu_14_PRA}%
  \BibitemOpen
  \bibfield  {author} {\bibinfo {author} {\bibfnamefont {G.}~\bibnamefont
  {Shu}}, \bibinfo {author} {\bibfnamefont {G.}~\bibnamefont {Vittorini}},
  \bibinfo {author} {\bibfnamefont {A.}~\bibnamefont {Buikema}}, \bibinfo
  {author} {\bibfnamefont {C.~S.}\ \bibnamefont {Nichols}}, \bibinfo {author}
  {\bibfnamefont {C.}~\bibnamefont {Volin}}, \bibinfo {author} {\bibfnamefont
  {D.}~\bibnamefont {Stick}}, \ and\ \bibinfo {author} {\bibfnamefont {K.~R.}\
  \bibnamefont {Brown}},\ }\href@noop {} {\bibfield  {journal} {\bibinfo
  {journal} {Phys. Rev. A}\ }\textbf {\bibinfo {volume} {89}},\ \bibinfo
  {pages} {062308} (\bibinfo {year} {2014})}\BibitemShut {NoStop}%
\bibitem [{\citenamefont {Olmschenk}\ \emph {et~al.}(2007)\citenamefont
  {Olmschenk}, \citenamefont {Younge}, \citenamefont {Moehring}, \citenamefont
  {Matsukevich}, \citenamefont {Maunz},\ and\ \citenamefont
  {Monroe}}]{Olmschenk_07_PRA}%
  \BibitemOpen
  \bibfield  {author} {\bibinfo {author} {\bibfnamefont {S.}~\bibnamefont
  {Olmschenk}}, \bibinfo {author} {\bibfnamefont {K.~C.}\ \bibnamefont
  {Younge}}, \bibinfo {author} {\bibfnamefont {D.~L.}\ \bibnamefont
  {Moehring}}, \bibinfo {author} {\bibfnamefont {D.~N.}\ \bibnamefont
  {Matsukevich}}, \bibinfo {author} {\bibfnamefont {P.}~\bibnamefont {Maunz}},
  \ and\ \bibinfo {author} {\bibfnamefont {C.}~\bibnamefont {Monroe}},\
  }\href@noop {} {\bibfield  {journal} {\bibinfo  {journal} {Phys. Rev. A}\
  }\textbf {\bibinfo {volume} {76}},\ \bibinfo {pages} {052314} (\bibinfo
  {year} {2007})}\BibitemShut {NoStop}%
\bibitem [{\citenamefont {Noek}\ \emph {et~al.}(2013)\citenamefont {Noek},
  \citenamefont {Vrijsen}, \citenamefont {Gaultney}, \citenamefont {Mount},
  \citenamefont {Kim}, \citenamefont {Maunz}, ,\ and\ \citenamefont
  {Kim}}]{Noek_2013_OL}%
  \BibitemOpen
  \bibfield  {author} {\bibinfo {author} {\bibfnamefont {R.}~\bibnamefont
  {Noek}}, \bibinfo {author} {\bibfnamefont {G.}~\bibnamefont {Vrijsen}},
  \bibinfo {author} {\bibfnamefont {D.}~\bibnamefont {Gaultney}}, \bibinfo
  {author} {\bibfnamefont {E.}~\bibnamefont {Mount}}, \bibinfo {author}
  {\bibfnamefont {T.}~\bibnamefont {Kim}}, \bibinfo {author} {\bibfnamefont
  {P.}~\bibnamefont {Maunz}}, , \ and\ \bibinfo {author} {\bibfnamefont
  {J.}~\bibnamefont {Kim}},\ }\href@noop {} {\bibfield  {journal} {\bibinfo
  {journal} {Optics Lett.}\ }\textbf {\bibinfo {volume} {38}},\ \bibinfo
  {pages} {4735} (\bibinfo {year} {2013})}\BibitemShut {NoStop}%
\bibitem [{Note1()}]{Note1}%
  \BibitemOpen
  \bibinfo {note} {In this evolution operator, we suppress the optical
  difference phase between the two beams generating the SDK because it is
  constant over the time scale of one experiment and ultimately
  cancels.}\BibitemShut {Stop}%
\bibitem [{\citenamefont {Monroe}\ \emph {et~al.}(1996)\citenamefont {Monroe},
  \citenamefont {Meekhof}, \citenamefont {King},\ and\ \citenamefont
  {Wineland}}]{Monroe_96_Science}%
  \BibitemOpen
  \bibfield  {author} {\bibinfo {author} {\bibfnamefont {C.}~\bibnamefont
  {Monroe}}, \bibinfo {author} {\bibfnamefont {D.}~\bibnamefont {Meekhof}},
  \bibinfo {author} {\bibfnamefont {B.}~\bibnamefont {King}}, \ and\ \bibinfo
  {author} {\bibfnamefont {D.}~\bibnamefont {Wineland}},\ }\href@noop {}
  {\bibfield  {journal} {\bibinfo  {journal} {Science}\ }\textbf {\bibinfo
  {volume} {272}},\ \bibinfo {pages} {1131} (\bibinfo {year}
  {1996})}\BibitemShut {NoStop}%
\bibitem [{\citenamefont {Sawyer}\ \emph {et~al.}(2014)\citenamefont {Sawyer},
  \citenamefont {Britton},\ and\ \citenamefont {Bollinger}}]{Sawyer_2014_PRA}%
  \BibitemOpen
  \bibfield  {author} {\bibinfo {author} {\bibfnamefont {B.~C.}\ \bibnamefont
  {Sawyer}}, \bibinfo {author} {\bibfnamefont {J.~W.}\ \bibnamefont {Britton}},
  \ and\ \bibinfo {author} {\bibfnamefont {J.~J.}\ \bibnamefont {Bollinger}},\
  }\href@noop {} {\bibfield  {journal} {\bibinfo  {journal} {Phys. Rev. A}\
  }\textbf {\bibinfo {volume} {89}},\ \bibinfo {pages} {033408} (\bibinfo
  {year} {2014})}\BibitemShut {NoStop}%
\bibitem [{\citenamefont {Glauber}(1963)}]{Glauber_1963_PR}%
  \BibitemOpen
  \bibfield  {author} {\bibinfo {author} {\bibfnamefont {R.~J.}\ \bibnamefont
  {Glauber}},\ }\href@noop {} {\bibfield  {journal} {\bibinfo  {journal} {Phys.
  Rev.}\ }\textbf {\bibinfo {volume} {131}},\ \bibinfo {pages} {2766} (\bibinfo
  {year} {1963})}\BibitemShut {NoStop}%
\bibitem [{\citenamefont {Sudarshan}(1963)}]{Sudarshan_1963_PRL}%
  \BibitemOpen
  \bibfield  {author} {\bibinfo {author} {\bibfnamefont {E.~C.~G.}\
  \bibnamefont {Sudarshan}},\ }\href@noop {} {\bibfield  {journal} {\bibinfo
  {journal} {Phys. Rev. Lett.}\ }\textbf {\bibinfo {volume} {10}},\ \bibinfo
  {pages} {277} (\bibinfo {year} {1963})}\BibitemShut {NoStop}%
\bibitem [{\citenamefont {Turchette}\ \emph
  {et~al.}(2000{\natexlab{b}})\citenamefont {Turchette}, \citenamefont {Myatt},
  \citenamefont {King}, \citenamefont {Sackett}, \citenamefont {Kielpinski},
  \citenamefont {Itano}, \citenamefont {Monroe},\ and\ \citenamefont
  {Wineland}}]{Turchette_00_PRA}%
  \BibitemOpen
  \bibfield  {author} {\bibinfo {author} {\bibfnamefont {Q.~A.}\ \bibnamefont
  {Turchette}}, \bibinfo {author} {\bibfnamefont {C.~J.}\ \bibnamefont
  {Myatt}}, \bibinfo {author} {\bibfnamefont {B.~E.}\ \bibnamefont {King}},
  \bibinfo {author} {\bibfnamefont {C.~A.}\ \bibnamefont {Sackett}}, \bibinfo
  {author} {\bibfnamefont {D.}~\bibnamefont {Kielpinski}}, \bibinfo {author}
  {\bibfnamefont {W.~M.}\ \bibnamefont {Itano}}, \bibinfo {author}
  {\bibfnamefont {C.}~\bibnamefont {Monroe}}, \ and\ \bibinfo {author}
  {\bibfnamefont {D.~J.}\ \bibnamefont {Wineland}},\ }\href@noop {} {\bibfield
  {journal} {\bibinfo  {journal} {Phys. Rev. A}\ }\textbf {\bibinfo {volume}
  {62}},\ \bibinfo {pages} {053807} (\bibinfo {year}
  {2000}{\natexlab{b}})}\BibitemShut {NoStop}%
\bibitem [{\citenamefont {Myatt}\ \emph {et~al.}(2000)\citenamefont {Myatt},
  \citenamefont {King}, \citenamefont {Turchette}, \citenamefont {Sackett},
  \citenamefont {Kielpinski}, \citenamefont {Itano}, \citenamefont {Monroe},\
  and\ \citenamefont {Wineland}}]{Myatt_00_Nature}%
  \BibitemOpen
  \bibfield  {author} {\bibinfo {author} {\bibfnamefont {C.~J.}\ \bibnamefont
  {Myatt}}, \bibinfo {author} {\bibfnamefont {B.~E.}\ \bibnamefont {King}},
  \bibinfo {author} {\bibfnamefont {Q.~A.}\ \bibnamefont {Turchette}}, \bibinfo
  {author} {\bibfnamefont {C.~A.}\ \bibnamefont {Sackett}}, \bibinfo {author}
  {\bibfnamefont {D.}~\bibnamefont {Kielpinski}}, \bibinfo {author}
  {\bibfnamefont {W.~M.}\ \bibnamefont {Itano}}, \bibinfo {author}
  {\bibfnamefont {C.}~\bibnamefont {Monroe}}, \ and\ \bibinfo {author}
  {\bibfnamefont {D.~J.}\ \bibnamefont {Wineland}},\ }\href@noop {} {\bibfield
  {journal} {\bibinfo  {journal} {Nature}\ }\textbf {\bibinfo {volume} {403}},\
  \bibinfo {pages} {269} (\bibinfo {year} {2000})}\BibitemShut {NoStop}%
\bibitem [{\citenamefont {Poschinger}\ \emph {et~al.}(2010)\citenamefont
  {Poschinger}, \citenamefont {Walther}, \citenamefont {Singer},\ and\
  \citenamefont {Schmidt-Kaler}}]{Poshinger_2010_PRL}%
  \BibitemOpen
  \bibfield  {author} {\bibinfo {author} {\bibfnamefont {U.}~\bibnamefont
  {Poschinger}}, \bibinfo {author} {\bibfnamefont {A.}~\bibnamefont {Walther}},
  \bibinfo {author} {\bibfnamefont {K.}~\bibnamefont {Singer}}, \ and\ \bibinfo
  {author} {\bibfnamefont {F.}~\bibnamefont {Schmidt-Kaler}},\ }\href@noop {}
  {\bibfield  {journal} {\bibinfo  {journal} {Phys. Rev. Lett.}\ }\textbf
  {\bibinfo {volume} {105}},\ \bibinfo {pages} {263602} (\bibinfo {year}
  {2010})}\BibitemShut {NoStop}%
\bibitem [{\citenamefont {Wineland}\ \emph {et~al.}(1998)\citenamefont
  {Wineland}, \citenamefont {Monroe}, \citenamefont {Itano}, \citenamefont
  {Leibfried}, \citenamefont {King},\ and\ \citenamefont
  {Meekof}}]{Wineland_1998_bible}%
  \BibitemOpen
  \bibfield  {author} {\bibinfo {author} {\bibfnamefont {D.~J.}\ \bibnamefont
  {Wineland}}, \bibinfo {author} {\bibfnamefont {C.}~\bibnamefont {Monroe}},
  \bibinfo {author} {\bibfnamefont {W.~M.}\ \bibnamefont {Itano}}, \bibinfo
  {author} {\bibfnamefont {D.}~\bibnamefont {Leibfried}}, \bibinfo {author}
  {\bibfnamefont {B.~E.}\ \bibnamefont {King}}, \ and\ \bibinfo {author}
  {\bibfnamefont {D.~M.}\ \bibnamefont {Meekof}},\ }\href@noop {} {\bibfield
  {journal} {\bibinfo  {journal} {J. Res. Natl. Inst. Stand. Technol.}\
  }\textbf {\bibinfo {volume} {103}},\ \bibinfo {pages} {259} (\bibinfo {year}
  {1998})}\BibitemShut {NoStop}%
\bibitem [{\citenamefont {Wineland}\ and\ \citenamefont
  {Itano}(1979)}]{Wineland_1979_PRA}%
  \BibitemOpen
  \bibfield  {author} {\bibinfo {author} {\bibfnamefont {D.~J.}\ \bibnamefont
  {Wineland}}\ and\ \bibinfo {author} {\bibfnamefont {W.~M.}\ \bibnamefont
  {Itano}},\ }\href@noop {} {\bibfield  {journal} {\bibinfo  {journal} {Phys.
  Rev. A}\ }\textbf {\bibinfo {volume} {20}},\ \bibinfo {pages} {1521}
  (\bibinfo {year} {1979})}\BibitemShut {NoStop}%
\bibitem [{\citenamefont {Gehrke}(2008)}]{Gehrke_thesis}%
  \BibitemOpen
  \bibfield  {author} {\bibinfo {author} {\bibfnamefont {C.}~\bibnamefont
  {Gehrke}},\ }\href@noop {} {Ph.D. thesis},\ \bibinfo  {school} {University of
  Rostock} (\bibinfo {year} {2008})\BibitemShut {NoStop}%
\bibitem [{\citenamefont {Wigner}(1932)}]{Wigner_1932_PR}%
  \BibitemOpen
  \bibfield  {author} {\bibinfo {author} {\bibfnamefont {E.}~\bibnamefont
  {Wigner}},\ }\href@noop {} {\bibfield  {journal} {\bibinfo  {journal} {Phys.
  Rev.}\ }\textbf {\bibinfo {volume} {40}},\ \bibinfo {pages} {749} (\bibinfo
  {year} {1932})}\BibitemShut {NoStop}%
\end{thebibliography}%
\end{document}